# Estimated Phase II Weibull control chart for monitoring times between events

Tanuja Negi

Department of Statistics, Faculty of Science, Banaras Hindu University, Varanasi, India

**Abstract**

In statistical process control (SPC), Weibull distribution can be used to model the time between events or failures (TBE) in a process with increasing, decreasing or constant failure rates. Specifically, it helps in monitor processes where the time between defect occurrences (or other relevant events) is of interest, providing insight into the process's reliability and performance. In this paper, we consider the two-sided problem of monitoring either an increase or a decrease in the scale parameter of the Weibull distribution with control charts assuming the shape parameter to be fixed. A larger scale parameter indicates that events are more spread out over time, suggesting fewer defects, while a smaller value may suggest deterioration in the process, resulting in a higher number of defects. When the scale parameter is unknown, it is estimated from Phase I observations and the plug-in control limits are obtained by replacing the parameter by its estimated value from the Phase I observations. Given the well-accepted fact that estimated control limits often do not perform as expected, the control limits of the Weibull chart are adjusted to achieve the desired in-control (IC) performance using two criteria: (i) conditional average run length (AARL), (ii) standard deviation of conditional ARL. A performance study is carried out in order to assess in-control and out-of-control performances of the proposed charts.

1. **Introduction**

In Industrial production, quality is one of the important decision making factors in the selection of any product and services. A good quality product satisfies the consumer needs specifying whether it is able to perform certain functions demanded by the customer. Such as in the airplanes, frequent need of repair demolishes the customer reliability and can take the life of people if not properly checked. The failure of the product increases the quality-related problem. For example, an automobile consists of thousands of parts. If every part of it is slightly defected, then this can lead to the vehicle not performing well as it is expected to perform. Moreover, any two products produced by the same process are not identical. There always exists an inevitable variation also known as inherent or natural variability. This background noise is often referred to as a stable state of chance. In the framework of statistical process monitoring, the variability may also occur occasionally and reason being improperly adjusted or controlled machines, error of operator, or defective raw materials. Such a variability is known as assignable variation.

To reduce this variation, Shewhart in 1924 developed the control chart to distinguish between this background noise and assignable variation. The control charts are tools for statistical process control and monitoring. Moreover, it is used to monitor the process output and detect the changes required to bring the process in-control state. In this research article we will study the Shewhart type Chart used to detect large shifts in a process, see Montgomery (2007).

It is a common practice to observe the time between the occurrence of two consecutive events or failure or defects also known as Time Between Events which is an important measure of failure rate of a product in order to ensure its reliability. And this failure rate is not always constant, so it is justifiable to model the

TBE using the Weibull distribution see Shafae et al (2015). The Weibull distribution has an increasing, decreasing and constant failure rate depending on the value of the shape parameter. Moreover, when the failure rate is constant the Weibull distribution is often associated with the amount of time taken for the occurrence of a random event such as time for the failure of hardware. Further, the decreasing failure rate is commonly perceived in semiconductors, and increasing failure rate in the components of mechanical systems, for instance, bearing wear see Montgomery (2010) and Rausand (2003). The shape parameter of the Weibull distribution provides critical information about the failure rate change over time whereas the scale parameter gives the variability present in the distribution. Furthermore, the scale parameter affects how far the probability distribution stretches out and its value is equivalent to 63.2 percentile in the distribution that is it gives 63.2% of the values in the distribution are less than the scale value. For instance, Yu et. al (2024) proposed three monitoring schemes to detect the shift in the scale parameter of Weibull distribution considering the progressively Type-II censored data. Li et. al introduced a one-sided control chart to monitor the downward mean shifts in TBE whose performance is further compared with other control charts.

It is notable that all these charts assumed parameters to be known (Case K). But, this situation rarely occurs in practice and it is very frequent to deal with the unknown parameters case (Case U). In this scenario, the plug-in control limits are constructed by replacing the estimator with parameters. This plug-in control chart performs differently from the expected control chart. Moreover, using estimated parameters for a long time in control charts originally designed under the assumption of known parameters, introduces extra variability that can considerably reduce their effectiveness. For instance, Jensen et al. (2006) discussed how parameter estimation affects the performance of various control charts. They specifically highlighted that this issue becomes more prominent when the Phase I sample size is small.

Raza et al. (2024) performed a comparative study on the mean charts assuming Weibull and generalized exponential distribution with both shape and scale parameters known. This further, concluded that a generalized exponential distribution based chart achieved the desired ARL value on a single value of the shape parameter whereas the Weibull distribution required a wide range of threshold values depending on its shape parameter. Furthermore, Silva et at. (2024) proposed a Shewhart $\bar{X}$ control chart to monitor the mean of the Discrete Weibull process using the Markov Chain procedure, which allowed the calculation of precise control limits. Furthermore, Ramalhoto et al. (2010) proposed Shewhart control chart for the scale parameter of Weibull with fixed and variable sampling intervals considering both the parameters known. Thus, to the best of our knowledge all the work dealt with the parameters known and there has been scarce work done in the unknown scale and shape parameter and the effect of the estimated parameters of Weibull distribution in the control chart.

The structure of this paper is as follows: The introductory section provides an overview of the Weibull distribution and the related Shewhart-type chart with probability limits, in two cases. Firstly, we will discuss Case I - when both the parameters are known and case II - when the shape parameter is known and scale is unknown. Also, a brief review on the comparsion of the performance measures in Case I with performance in Case II is done. The simulation study on the performance of the considered chart is performed.

2. **Control Chart for assessing Time Between Events of Weibull Distribution**

This section briefly discusses the Weibull distribution with the two-sided Shewhart chart in the case of both known parameters denoted by Case I and Case II of shape parameter known and scale unknown.

## The Weibull Distribution

Let X be a random variable that follow Weibull distribution with scale parameter β and shape parameter η. Then, its probability density function is given as:

$$f(x; \eta, \beta) = \frac{\eta}{\beta}\left(\frac{x}{\beta}\right)^{\eta-1} e^{-\left(\frac{x}{\beta}\right)^{\eta}}, \quad x > 0, \beta > 0, \eta > 0 \quad [1]$$

The expected value and the variance of X are given by:

$$E(x) = \beta \, \Gamma(1 + 1/\eta)$$

$$V(x) = \beta^2 \left[\Gamma\left(1 + \frac{2}{\eta}\right) - \left(\Gamma\left(1 + \frac{1}{\eta}\right)\right)^2\right]$$

Also, the distribution function and the inverse distribution are, respectively, given by:

$$F(x; \eta, \beta) = P(X \leq x) = \int_0^\infty f(x; \eta, \beta) dx = 1 - e^{-\left(\frac{x}{\beta}\right)^{\eta}}, \quad x > 0 \quad [2]$$

$$F^{-1}(u; \eta, \beta) = \beta[1 - \ln(1 - u)]^{\frac{1}{\eta}}, \quad 0 < u < 1 \quad [3]$$

where $F^{-1}(u; \eta, \beta)$ is the $u$-percentile point of the Weibull distribution.

## Case 1: The Shewhart chart with both known parameters

Let $\alpha_0$ is the False Alarm Rate which is the probability of giving signal when the process is in-control and the shape and scale parameter be $\eta_0$ and $\beta_0$ respectively. Further, in Phase II monitoring let both the parameters be shifted to $\eta_1$ and $\beta_1$.

The lower and the upper control limits $LCL_K$, $UCL_K$ are:

$$P[X < LCL_K] = \frac{\alpha_0}{2} \text{ and } P[X > UCL_K] = \frac{\alpha_0}{2}$$

Where the in-control values of the shape and scale parameter are $\eta_0$ and $\beta_0$ and $X \sim W(\eta_0, \beta_0)$.

Thus, the control limits are

$$P[X < LCL_K] = \frac{\alpha_0}{2} \Rightarrow LCL_K = \beta . A_1 \text{ where } A_1 = \left[-\ln\left(1 - \frac{\alpha_0}{2}\right)\right]^{\frac{1}{\eta_0}}, \quad [4]$$

$$P[X > UCL_K] = \frac{\alpha_0}{2} \Rightarrow UCL_K = \beta . A_2 \text{ where } A_2 = \left[-\ln\left(\frac{\alpha_0}{2}\right)\right]^{\frac{1}{\eta_0}} \quad [5]$$

And, the central line CL is:

$$CL_K = \beta_0 \Gamma(1 + 1/\eta_0) \quad [6]$$

Let $\delta_1 = \frac{\beta_1}{\beta_0}$ and $\delta_2 = \frac{\eta_1}{\eta_0}$ specify the amount of shift in the scale and shape parameter from $\beta = \beta_0$ (in-control) to some value of $\beta = \beta_1$ (out-of-control) and $\eta = \eta_0$ (in-control) to some value of $\eta = \eta_1$ (out-of-control) respectively. Then, the probability to signal (PS) is the probability that the plotting statistic gives a signal, that is, it falls beyond the control limits. This is given by:

$$PS = P(X < LCL_K \text{ or } X > UCL_K)$$
$$= 1 - exp\left(-\left(\frac{A_1}{\delta_1^{\eta}}\right)^{\delta_2}\right) + exp\left(-\left(\frac{A_2}{\delta_1^{\eta}}\right)^{\delta_2}\right) \qquad [7]$$

The number of points plotted before a point indicates an out of control signal condition is called the run length and for the Shewhart chart it follows geometric distribution. Thus, the Run Length (ARL) for the Weibull chart with control limits ($LCL_K$, $UCL_K$) will follow Geometric distribution with parameter PS. The, the Average Run Length (ARL) is given by:

$$ARL = \frac{1}{1 - exp\left(-\left(\frac{A_1}{\delta_1^{\eta}}\right)^{\delta_2}\right) + exp\left(-\left(\frac{A_2}{\delta_1^{\eta}}\right)^{\delta_2}\right)} \qquad [8]$$

And, for the IC ARL value, that is $ARL_{in}$, is obtained by substituting $\delta_1 = \delta_2 = 1$.

$$ARL_{in} = ARL(1) = \frac{1}{1 - exp(-A_1) + exp(A_2)} = \frac{1}{FAR}$$

This is similar to the result obtained by Kumar et al. (2025). And depends only on the False Alarm Rate ($\alpha_0$).

**Case II: The Shewhart chart with shape parameter known and scale parameter unknown**

In this case the value of the shape parameter is known i.e. the failure rate of the observation is known and the scale parameter is not known, then this must be estimated from the Phase I sample with the use of Maximum Likelihood Estimation and substituting this estimated value of the scale parameter plug in control limits are obtained. Let $X_1, X_2, \ldots, X_m$ be a phase I sample of size m from the $W(\eta_0, \beta_0)$ distribution. For given values $x_1, x_2, \ldots, x_m$, the log-likelihood function is

$$l(\eta, \beta) = -m\eta(ln\beta) - \left(\frac{\sum_{i=1}^{m} x_i}{\beta^{\eta}}\right)$$

Further, maximizing the above function for scale parameter $\beta$. The MLE estimate of $\beta$ is

$$\hat{\beta} = \left(\frac{\sum_{i=1}^{m} x_i^{\eta}}{m}\right)^{\frac{1}{\eta}}$$

Similar to Case K, the control limits $LCL, UCL$ are:

$$P[X < LCL_K] = \frac{\alpha_0}{2} \Rightarrow LCL = \hat{\beta} A_1 \quad \text{where } A_1 = \left[-ln\left(1 - \frac{\alpha_0}{2}\right)\right]^{\frac{1}{\eta_0}}, \qquad [9]$$

$$P[X > UCL_K] = \frac{\alpha_0}{2} \Rightarrow UCL_K = \hat{\beta}.A_2 \quad \text{where } A_2 = \left[-\ln\left(\frac{\alpha_0}{2}\right)\right]^{\frac{1}{\eta_0}} \quad [10]$$

The control limits are known as conditional control limits that are conditioned on a given phase I sample. Now, to estimate the conditional performance of the estimated control chart, we consider the CRL conditioned on a given phase I sample which follow geometric distribution with parameter

$$CPS = P(X < LCL | \beta_1) + P(X > UCL | \beta_1)$$

$$= 1 - \exp\left\{-\left(\frac{\hat{\beta}}{\delta_1 * \beta_0}\right)^{\delta_2 \eta} A_1^{(\delta_2)}\right\} + \exp\left\{-\left(\frac{\hat{\beta}}{\delta_1 * \beta_0}\right)^{\delta_2 \eta} A_2^{(\delta_2)}\right\} \quad [11]$$

The average of the Conditional Run Length is:

$$CARL = \frac{1}{CPS} = \frac{1}{1-\exp\left\{-\left(\frac{\hat{\beta}}{\delta_1 * \beta_0}\right)^{\delta_2 \eta} A_1^{(\delta_2)}\right\}+\exp\left\{-\left(\frac{\hat{\beta}}{\delta_1 * \beta_0}\right)^{\delta_2 \eta} A_2^{(\delta_2)}\right\}} \quad [12]$$

Now, considering the In-Control process, that is $\delta_1 = 1$ and $\delta_2 = 1$, the CARL is reduced to

$$CARL = \frac{1}{CPS} = \frac{1}{1-\exp\left\{-\left(\frac{\hat{\beta}}{\beta_0}\right)^{\eta} A_1\right\}+\exp\left\{-\left(\frac{\hat{\beta}}{\beta_0}\right)^{\eta} A_2\right\}}$$

Here, the distribution of $W = \left(\frac{\hat{\beta}}{\beta_0}\right)^{\eta}$ is Gamma distribution with shape and scale parameter $m$ and $m$ respectively. Let $Y = 2wm$, this follows chi square distribution with degree of freedom $2m$. Then, the expected CARL is evaluated as:

$$ECARL = \int_0^\infty (CARL) f(y) d(y) = \int_0^\infty \frac{1}{1-\exp\left\{-\frac{YA_1}{2m}\right\}+\exp\left\{-\frac{YA_2}{2m}\right\}} f(y) dy$$

And the standard deviation of CARL is evaluated as:

$$SDCARL = E(CARL)^2 - (E(CARL))^2$$

### 3. **Numerical Results**

#### 3.1. **Performance of the Shewhart chart for Case II**

This part of study investigates the effect of estimated scale parameters on the performance of CARL of the Shewhart chart. The simulation study is done by constructing the tables for different values of $m$ and parameters.

In Table 1, the results are obtained by simulation for $\alpha = 0.0027$ resulting in the nominal IC ARL of 370.4 and $m \in \{30, 50, 100, 200, 500, 800, 1000, 2000, 5000, 8000\}$. The results are tabulated for the AARL and SDARL with the $\gamma$ percentile points that gives the information on the variability of CARL with $\gamma \in \{0.05, 0.10, 0.25, 0.50, 0.75, 0.90, 0.95\}$. An Exceedance probability defined as the probability of CARL to be less than nominal ARL value 370.4 is tabulated in column EPC.

In the Table 1 it is clearly shown that as $m$ increases the value of ACARL converges to the nominal ARL of 370.4 and the standard deviation of CARL reduces which is the expected result. Moreover, the value of exceedance probability (EPC) at least 50% that is at least 50% of the CARL value will be less than 370.4 and the control chart will signal frequently during the in-control conditions resulting in increased False

Alarm Rate (FAR). This will lead to deviation from IC performance. Also, it can seen that for $m > 5000$ The ACARL is close to the In-Control ARL and the standard deviation is reduced. Moreover, the ACARL is smaller than the IC-ARL for small values of $m$.

**Table 1: The distribution of the CARL when control limits are used with shape parameter known**

| $(\eta_0, \beta_0)$ | m | ACARL | SDCARL | EPC | 5% | 10% | 25% | 50% | 75% | 90% | 95% |
|---|---|---|---|---|---|---|---|---|---|---|---|
| (1, 1) | 30 | 338.9346 | 135.7551 | 52.925% | 102.2866 | 139.3152 | 227.4791 | 357.158 | 465.3308 | 506.8789 | 513.4964 |
| | 50 | 351.6829 | 115.8351 | 51.49% | 145.7871 | 183.678 | 263.5128 | 364.813 | 453.2411 | 499.0031 | 510.5038 |
| | 100 | 359.6624 | 90.08362 | 51.775% | 200.8644 | 235.0289 | 295.1394 | 365.82 | 431.609 | 476.7128 | 494.5342 |
| | 200 | 365.3853 | 68.32933 | 50.506% | 246.167 | 272.2146 | 319.2629 | 369.462 | 416.8051 | 452.2679 | 470.6770 |
| | 500 | 367.2168 | 44.63903 | 51.22% | 290.8276 | 307.5175 | 336.9271 | 369.141 | 398.6793 | 423.9756 | 437.9931 |
| | 800 | 369.1364 | 35.69011 | 50.5% | 308.6365 | 322.7094 | 345.0873 | 369.897 | 394.2097 | 414.8554 | 426.3074 |
| | 1000 | 369.7356 | 32.42453 | 49.95% | 315.0153 | 327.3117 | 347.8848 | 370.432 | 392.3045 | 411.2402 | 421.9334 |
| | 2000 | 369.7555 | 22.85858 | 50.55% | 331.577 | 339.8705 | 354.2959 | 370.099 | 385.449 | 398.9634 | 406.764 |
| | 5000 | 370.0579 | 14.66042 | 50.555% | 345.8348 | 351.143 | 360.1586 | 370.2 | 379.9169 | 388.8192 | 393.9695 |
| | 8000 | 370.311 | 11.66266 | 49.745% | 350.9131 | 355.3566 | 362.4859 | 370.481 | 378.3144 | 385.1541 | 389.2243 |
| (1, 15) | 30 | 341.4732 | 136.2757 | 51.76% | 101.92 | 139.58 | 230.34 | 361.04 | 469.03 | 507.99 | 513.84 |
| | 50 | 348.0246 | 116.6252 | 52.93% | 143.44 | 179.19 | 258.22 | 359.49 | 449.32 | 498.09 | 510.51 |
| | 100 | 360.045 | 90.70242 | 51.135% | 201.66 | 233.21 | 294.46 | 367.63 | 431.78 | 477.15 | 495.39 |
| | 200 | 365.2565 | 68 | 50.965% | 246.53 | 273.27 | 319.09 | 368.43 | 416.41 | 452.37 | 471.04 |
| | 500 | 367.3603 | 45.0659 | 51.22% | 291.2 | 307.98 | 336.98 | 369.19 | 399.07 | 424.52 | 438.8 |
| | 800 | 369.1633 | 36.10612 | 50.15% | 307.17 | 321.69 | 345.28 | 370.24 | 394.58 | 414.57 | 426.7 |
| | 1000 | 369.0211 | 32.2744 | 50.63% | 314.75 | 327.15 | 347.25 | 369.91 | 391.27 | 410.27 | 421.26 |
| | 2000 | 369.9268 | 23.14387 | 50.1% | 331.04 | 339.88 | 354.34 | 370.35 | 386.24 | 399.52 | 406.85 |
| | 5000 | 370.1558 | 14.66352 | 50.42% | 345.86 | 351.39 | 360.31 | 370.26 | 380.14 | 388.96 | 394.15 |
| | 8000 | 370.2034 | 11.62036 | 50.51% | 350.82 | 355.42 | 362.39 | 370.24 | 378.14 | 384.97 | 389.08 |
| (0.5,10) | 30 | 340.5756 | 135.1953 | 52.51% | 105.34 | 142.19 | 230.32 | 357.61 | 467.11 | 507.12 | 513.31 |
| | 50 | 350.8742 | 116.7905 | 51.85% | 146.76 | 183.61 | 260.65 | 363.48 | 454.52 | 500.55 | 511.45 |
| | 100 | 359.9434 | 90.28034 | 51.59% | 201.78 | 233.64 | 295.87 | 366.26 | 431.42 | 477.1 | 495.15 |
| | 200 | 365.4355 | 68.29744 | 50.335% | 246.51 | 272.48 | 317.83 | 369.9 | 416.73 | 452.68 | 470.83 |
| | 500 | 367.7202 | 44.67552 | 51% | 291.58 | 308.27 | 337.41 | 369.27 | 399.73 | 424.9 | 438.4 |
| | 800 | 368.9693 | 35.70416 | 50.975% | 309.71 | 322.39 | 344.57 | 369.54 | 394.01 | 415.01 | 426.48 |

| | | | | | | | | | | |
|---|---|---|---|---|---|---|---|---|---|---|
| | 1000 | 369.2059 | 32.38762 | 50.42% | 314.78 | 327.22 | 347.46 | 370.07 | 391.95 | 410.23 | 421.14 |
| | 2000 | 369.6665 | 23.01293 | 50.465% | 331.09 | 339.86 | 354.17 | 370.16 | 385.73 | 398.76 | 406.81 |
| | 5000 | 370.1484 | 14.50118 | 50.01% | 345.98 | 351.35 | 360.43 | 370.39 | 380.02 | 388.53 | 393.88 |
| | 8000 | 370.3285 | 11.55827 | 49.925% | 351.16 | 355.49 | 362.51 | 370.43 | 378.32 | 385.05 | 389.22 |
| (1.5 , 5) | 30 | 340.3942 | 135.6623 | 52.165% | 104.59 | 142.31 | 227.21 | 360 | 467.21 | 507.01 | 513.47 |
| | 50 | 351.3275 | 116.1451 | 51.89% | 143.93 | 182.67 | 262.74 | 363.33 | 452.82 | 499.71 | 511.13 |
| | 100 | 359.9385 | 90.62933 | 51.375% | 201.29 | 232.66 | 295.71 | 366.78 | 432.17 | 477.03 | 495.76 |
| | 200 | 365.1172 | 68.21157 | 50.825% | 247.39 | 271.97 | 318.52 | 369 | 416.26 | 453.13 | 471.27 |
| | 500 | 368.2431 | 44.84934 | 50.695% | 291.52 | 309.36 | 338 | 369.66 | 399.97 | 425.59 | 439.26 |
| | 800 | 369.0526 | 35.91992 | 50.015% | 308.45 | 321.45 | 344.81 | 370.38 | 394.47 | 414.36 | 426.54 |
| | 1000 | 369.4935 | 332.4601 | 49.765% | 314.5 | 326.93 | 347.59 | 370.6 | 391.79 | 410.91 | 421.91 |
| | 2000 | 370.0321 | 23.16496 | 49.89% | 331.42 | 339.9 | 354.55 | 370.47 | 386.12 | 399.42 | 407.37 |
| | 5000 | 370.0701 | 14.60575 | 50.165% | 345.75 | 351.17 | 360.28 | 370.34 | 380.06 | 388.59 | 393.69 |
| | 8000 | 370.2749 | 11.56566 | 50.04% | 350.83 | 355.35 | 362.58 | 370.39 | 378.1 | 385.11 | 389.23 |
| (2 , 5) | 30 | 340.5813 | 135.2365 | 52.495% | 106.13 | 143.21 | 229.56 | 359.15 | 467.85 | 507.26 | 513.28 |
| | 50 | 350.4566 | 116.485 | 52.015% | 143.68 | 182.15 | 261.62 | 363.1 | 450.79 | 500.27 | 511.09 |
| | 100 | 358.8118 | 91.00969 | 51.725% | 199.18 | 232.1 | 293.02 | 365.35 | 431.63 | 476.41 | 495.07 |
| | 200 | 363.6693 | 68.39092 | 51.61% | 245.35 | 270.47 | 316.66 | 367.39 | 415.16 | 451.49 | 469.97 |
| | 500 | 369.3343 | 44.68141 | 49.735% | 293.07 | 309.95 | 339.3 | 370.75 | 401.27 | 426.58 | 440.83 |
| | 800 | 368.0775 | 36.24358 | 50.65% | 307.86 | 321.62 | 344.74 | 369.87 | 394.29 | 415.67 | 427.33 |
| | 1000 | 369.5094 | 32.279 | 50.06% | 314.67 | 327.32 | 347.53 | 370.36 | 392.11 | 410.94 | 421.78 |
| | 2000 | 369.8414 | 23.00679 | 50.65% | 331.51 | 340.32 | 354.22 | 369.99 | 385.69 | 399.36 | 407 |
| | 5000 | 370.2268 | 14.6766 | 49.85% | 345.68 | 351.2 | 360.41 | 370.46 | 380.2 | 388.89 | 394.01 |
| | 8000 | 370.2304 | 11.54041 | 50.3% | 350.92 | 355.4 | 362.48 | 370.32 | 378.16 | 385.04 | 388.98 |
| (10 ,15 ) | 30 | 570.1816 | 1163.433 | 63.425% | 38.2 | 55.42 | 109.24 | 241.92 | 567.68 | 1282.41 | 2110.79 |
| | 50 | 487.3958 | 699.6413 | 60.32% | 63.87 | 85.87 | 150.68 | 285.26 | 565.07 | 1049.33 | 1527.13 |
| | 100 | 422.2673 | 350.9902 | 57.285% | 107.38 | 136.18 | 204.82 | 325.79 | 523.46 | 808.23 | 1047.17 |
| | 200 | 394.6308 | 211.3515 | 55.255% | 155.23 | 184.62 | 247.51 | 347.23 | 486.54 | 662.33 | 794.45 |
| | 500 | 380.3654 | 124.7512 | 53.35% | 215.06 | 241.17 | 291.44 | 360.59 | 449 | 545.45 | 608.95 |
| | 800 | 375.8945 | 95.95165 | 53.09% | 242.02 | 264.32 | 307.63 | 363.57 | 432.11 | 501.86 | 549.18 |
| | 1000 | 376.1986 | 86.33391 | 52.32% | 254.76 | 275.53 | 314.88 | 365.78 | 426.41 | 490.37 | 535.33 |

| | 2000 | 372.5184 | 59.07884 | 52.01% | 284.08 | 300.9 | 330.71 | 367.5 | 409.57 | 451.04 | 477.04 |
| | 5000 | 371.1938 | 37.44098 | 51.105% | 312.94 | 324.58 | 345.17 | 369.38 | 395.1 | 419.56 | 435.98 |
| | 8000 | 370.7668 | 29.47538 | 51.325% | 323.65 | 333.82 | 350.42 | 369.41 | 390.36 | 408.97 | 421.36 |

## 4. Conclusion

In this paper, we investigated the effects of parameter estimation of scale parameter on the ARL and CARL of the two-sided control charts with estimated control limits. Further, under the both parameter known case, the ARL value is 370.4 for all values of phase I sample whereas in the case of shape parameter known and scale unknown, the ARL values converges to the nominal value and will increase the value of size of phase I sample.

**Reference**

Shafae MS , Dickinson RM , Woodall WH , Camelio JA .Cumulative sum control charts for monitoring Weibull-distributed time between events.Quality and Reliability Engineering International 2015 ; 31(5) : 839-849.

Jensen, W.; Jones-Farmer, L. A.; Champ, C. W.; andWoodall, W. H. "Effects of Parameter Estimation on Control Chart Properties". Journal of Quality Technology 2006 ; 38(4), 349–364.

Montgomery DC. Introduction to Statistical Quality Control. Wiley & Sons; 2007.

Montgomery DC , RungerGC . Applied Statistics and Probability for Engineers. 2010.

Rausand M, Hoyland A. System Reliability Theory: Models, Statistical Methods, and Applications. 2003 ; 396.

Yu J., Wange Q.i, Wu C. Online monitoring of the Weibull distributed process based on progressive Type II censoring scheme. Journal of Computational and Applied Mathematics 2024 ; 443(16).

Raza A., Ali S., Shah I., Al-Rezami A.Y., Almazah M.M.A. A comparative analysis of mean charts assuming Weibull and generalized exponential distributions. Heliyon 2024 ; 10(21).

Silva L. A., Ho. L.L., Quinino R.D.C. Markov Chain approach to get control limits for a Shewhart Control Chart to monitor the mean of Discrete Weibull distribution. Journal of Process Control 2024 ; 134.

Jensen W.A., Farmer L.A.J., Champ C.W., Woodall W.H. Effect of Parameter Estimation on Control Chart Properties: A Literature Review 2018 ; 38(4) ; 349-364.

Ramalhoto M.F., Morais M., Shewhart control charts for the scale parameter of a Weibull control variable with fixed and variable sampling intervals 2010 ; 26(1) ; 129-160.